\documentclass{elsart}

\usepackage {amsfonts}
\usepackage {graphicx}
\usepackage {epsfig}
\usepackage{amssymb}
\usepackage{amsmath}
\usepackage {longtable}
\usepackage[english]{babel}
\begin{document}

\begin{frontmatter}
\title{Spin rotation and oscillations for high energy particles in a crystal and
possibility to measure the quadrupole moments and tensor
polarizabilities of elementary particles and nuclei}
\author{V.G. Baryshevsky}
\author{A.A. Gurinovich}

\address{Research Institute for Nuclear Problems, Belarusian State University,
11 Bobruiskaia Str., Minsk, 220050, Belarus}

\begin{abstract}
It is shown that particle motion in a bent (straight) crystal is
accompanied by particle spin rotation and oscillations that allows
to measure the tensor electric and magnetic polarizabilities of
nuclei and elementary particles. It is shown that channelling of
particles in either straight or bent crystal with the polarized
nuclei could be used both to analyze polarization of high energy
particles and polarize them.
\end{abstract}


\end{frontmatter}

\newpage

 \section{INTRODUCTION}
A new method for  high energy charged particle beams guiding by
bent crystals has been developing for more than 20 years in
several prominent laboratories in the world.
 The idea to use bent crystals for particle beam extraction from a
 storage ring was proposed in 1976
 \cite{tsyganov}. Now this idea is experimentally tested and
 becomes a basis for the method for particle beam extraction from a
 storage ring with sufficiently high efficiency ($ \sim 85 \%$).

 The next step in application of crystals for  high energy
 particle beams guiding was done in 1979 in \cite{VG79},
 where it was first shown that the spin of a high energy particle moving in
 a  bent crystal rotates with respect to its momentum due to
 the particle anomalous moment (see also \cite{n3}).
 Nowadays this effect has been experimentally
 discovered \cite{n4,n4a,n4b}.
  It has been also shown that the similar experiments are beneficial for
  measuring the magnetic moments of short-lived particles \cite{nn7}
  and, particularly, $\Lambda_c$ particle \cite{nn8}.
 Bent crystals can be used to steer particle
 spin orientation, too.
  Moreover, according to \cite{n5,n6} channelling of high energy
  particles in a crystal can be used for measuring the quadrupole
  moments of elementary particles, $\Omega^-$ hyperon in
  particular.

  The present paper shows that study of relativistic particle (nucleus) spin behavior in
  crystals (either bent or straight) allows to find the tensor
  electric and magnetic polarizabilities of the particle (nucleus).
  It is shown that phenomenon of channelling in either straight or bent crystal with the polarized nuclei
  can
be used either for making high energy particle beams polarized or
for polarization analysis.

  \section{The effective periodic potential of a crystal}

A fast particle passing through a monocrystal is elastically and
inelastically scattered due to interaction with electrons and
nuclei.
The secondary waves appear in the crystal due to particle
scattering.
It is important that the secondary waves describing elastic
scattering (i.e. scattering without crystal excitation) interfere
with each other and with the incident wave producing the sum
coherent wave in the crystal. The effective periodic potential
$U(\vec{r})$ can be introduced to describe passing of the coherent
wave in the crystal:
\begin{eqnarray}
U(\vec{r})=\sum_{\vec{\tau}} U(\vec{\tau}) e^{i \vec{\tau}
\vec{r}},
\end{eqnarray}
where $\vec{\tau}$ is the reciprocal lattice vector of the
crystal,
\begin{eqnarray}
U(\vec{\tau})=\frac{1}{V} \sum_{j} U_{j0}(\vec{\tau})
 e^{-W_j (\vec{\tau})} e^{i \vec{\tau} \vec{r}_j}
\end{eqnarray}
here $V$ is the volume of the crystal elementary cell, $\vec{r}_j$
is the coordinate of an atom (nucleus) of type $j$ in the crystal
elementary cell and the squared $e^{-W_j (\vec{\tau})}$ is equal
to the thermal-factor (i.e. Debye-Waller factor) well-known for
X-ray scattering \cite{James},
\begin{eqnarray}
U_{j0}(\vec{\tau}) =-\frac{2 \pi \hbar^2}{M \gamma}
F_j{(\vec{\tau})}
\end{eqnarray}
$M$ is the mass of the incident particle, $\gamma$ is its
Lorentz-factor, $F_j{(\vec{\tau})}=F_j{(\vec{k}^\prime-\vec{k}
=\vec{\tau})}$ is the amplitude of elastic coherent scattering of
the particle by the atom, $\vec{k}$ is the wave-vector of the
incident wave and  $\vec{k}^\prime$ is the wave vector of the
scattered wave.

Elastic coherent scattering of a particle by an atom is caused by
both Coulomb interaction of the particle with the atom electrons
and nucleus and its nuclear interaction with the nucleus.
Therefore, the scattering amplitude can be presented as a sum of
two amplitudes:
\begin{eqnarray}
F_j{(\vec{\tau})}=F_j^{coul}{(\vec{\tau})}+F_j^{nucl}{(\vec{\tau})}
\end{eqnarray}
where $F_j^{coul}{(\vec{\tau})}$ is the amplitude of particle
scattering caused by Coulomb interaction with the atom (nucleus)
(it contains contributions from the Coulomb interaction of the
particle with the atom along with the spin-orbit interaction with
the Coulomb field of the atom (nucleus));
$F_j^{nucl}{(\vec{\tau})}$ is the amplitude of elastic coherent
scattering of the particle caused by nuclear interaction
(this amplitude contains terms independent on the incident
particle spin along with terms depending on spin of both the
incident particle and nucleus, in particular, spin-orbit
interaction). Therefore, $U(\vec{r})$ and $U(\vec{\tau})$ also can
be expressed:
\begin{eqnarray}
\begin{array}{l}
U{(\vec{r})}=U^{coul}{(\vec{r})}+U^{nucl}{(\vec{r})},
\\
U{(\vec{\tau})}=U^{coul}{(\vec{\tau})}+U^{nucl}{(\vec{\tau})}
\end{array}
\end{eqnarray}
Suppose a high energy particle moves in a crystal at a small angle
to the crystallographic planes (axes) close to the Lindhard angle
$\vartheta_L \sim \sqrt{\frac{U}{E}}$ (in relativistic case
$\vartheta_L \sim \sqrt{\frac{2U}{E}}$), where $E$ is the energy
of the particle, $U$ is the height of the potential barrier
created by the crystallographic plane (axis).
This motion is determined by the plane (axis) potential $\hat{U}
(\vec{\rho})$, which could be derived from ${U} (\vec{r})$ by
averaging over the distribution of atoms (nuclei) in the crystal
plane (axis).

As a consequence, the potential $\hat{U} (\vec{\rho})$ for a
particle channelled in a plane (or axis) channel or moving over
the  barrier at a small angle close to the Lindhard angle can be
expressed as a sum:
\begin{eqnarray}
\hat{U}
(\vec{\rho})=\hat{U}^{coul}{(\vec{\rho})}+\hat{U}^{sp.-orb.}{(\vec{\rho})}+\hat{U}_{eff}^{nucl}{(\vec{\rho})}+
\hat{U}^{magn}{(\vec{\rho})}
\end{eqnarray}
where $\hat{U}^{coul}$ is the potential energy of particle Coulomb
interaction with the crystallographic plane (axis),
$\hat{U}^{sp.-orb.}$ is the energy of spin-orbit interaction with
the Coulomb field of the plane (axis) and spin-orbit nuclear
interaction with the effective nuclear field of the plane (axis),
$\hat{U}_{eff}^{nucl}$ is the effective potential energy of
nuclear interaction of the incident particle with the
crystallographic plane (axis)
\begin{eqnarray}
\hat{U}_{eff}^{nucl}{(\vec{\rho})}=-\frac{2 \pi \hbar^2}{M \gamma}
N{(\vec{\rho})} \hat{f}(0),
\end{eqnarray}
where $N{(\vec{\rho})}$ is the nuclei density in the point
$\vec{\rho}$ of the crystallographic plane (axis),
$\vec{\rho}=(x,y)$, the axis $z$ is directed along the crystal
axis (along the velocity component parallel to the
crystallographic plane in the case of plane channelling),
$\hat{f}(0)$ is the amplitude of elastic coherent forward
scattering, which depends on the incident particle spin and
nucleus polarization,
$\hat{U}^{magn}{(\vec{\rho})}$ is the energy of magnetic
interaction of the particle with the magnetic field created by
electrons (nuclei).

Thus, several contributions to the energy of interaction should be
considered for a high-energy particle elastically scattered in a
crystal.

Let us consider first the influence of the electromagnetic field
induced by the crystallographic planes (axes) on the particle
spin.

  \section{Spin rotation of relativistic particles in a crystal}

 Thus, let a high-energy particle travels through a
crystal.
 Inasmuch as in this case the particle wavelength is much
less then typical inhomogeneities in the electric fields in
crystals, so to describe spin motion in these fields a
quasi-classical approximation can be used.
  The equations describing particle spin motion in an   electromagnetic
field were obtained by Bargmann et. al. \cite{Bargmann}.
\begin{eqnarray}
\frac{d \vec{P}}{d t} = \left[ \vec{\Omega} \vec{P} \right],
 \label{0.1}
\end{eqnarray}
where $\vec{P}$ is the vector polarization of the particle,
\begin{eqnarray}
\Omega  =  -\frac{e}{M c} \left\{
\left(\frac{g-2}{2}+\frac{1}{\gamma}\right)\vec{B} -
\frac{g-2}{2}~ \frac{\gamma-1}{\gamma} \frac{(\vec{{v}}
\vec{B})\vec{v}}{v^2} -
\left(\frac{g-2}{2}+\frac{1}{\gamma+1}\right) \left[ \vec{\beta}
\times \vec{E} \right] \right\}\, \nonumber
\end{eqnarray}
is the angular velocity of particle spin precession at the
particle location point at time moment $t$,
$\vec{\beta}=\frac{\vec{v}}{c}$, $\vec{v}$ is the particle
velocity, $e$ is the particle charge, $M$ is its mass, $c$ is the
speed of light, $\gamma$ is the particle Lorentz factor, $g$ is
the gyromagnetic ratio ($g$-factor). In the absence of magnetic
field the angular velocity of particle momentum rotation at the
moment t is:
\begin{eqnarray}
\Omega_0 = \left[ \vec{n} \frac{d \vec{n}}{dt} \right]= -
\frac{e}{M c} \frac{\gamma}{\gamma^2-1}\left[\vec{E} \times
\vec{\beta}  \right], \nonumber
\end{eqnarray}
where $\vec{n}=\vec{v}/v$.

If $\gamma \gg 1$
\begin{eqnarray}
  \Omega_0 =
 - \frac{e}{M c \gamma}
\left[\vec{E} \times \vec{\beta} \right]=- \frac{e}{M c \gamma}
\left[\vec{E} \times \vec{n} \right] \nonumber
\end{eqnarray}
and
\begin{eqnarray}
\Omega  = \left(\frac{g-2}{2}+\frac{1}{\gamma}\right) \gamma
\Omega_0.  \nonumber
\end{eqnarray}
We should also recall that the anomalous magnetic moment of a
particle $\mu^{\prime}$ is in the following relation to (g-2):
\begin{eqnarray}
\mu^{\prime}=\frac{e (g-2)}{2 M c}\hbar S, \nonumber
\end{eqnarray}
where $S$ is the particle spin.

Let us consider first the simplest case for the sake of more clear
understanding. Suppose a particle move in a planar channelling
mode in the plane
 $x$, $z$ in a crystal bent about the $y$-axis.
  In this case from (\ref{0.1}) it follows that
\begin{eqnarray}
 \frac{dP_x}{dt}=\Omega_y(t)P_z,~
 \frac{dP_z}{dt}=\Omega_y(t)P_x,~
 \frac{dP_y}{dt}=0,
 \label{0.2}
\end{eqnarray}
where the frequency of particle precession  about the $y$-axis is:
\begin{eqnarray}
 \Omega_y(t)  = \left(\frac{g-2}{2}+\frac{1}{\gamma}\right) \gamma
\Omega_{0y}(t), \\
 |\Omega_y|=\left(\frac{g-2}{2}+\frac{1}{\gamma}\right) \frac{e}{M c}
E(t), \nonumber
 \label{0.3}
\end{eqnarray}
here $E$ is the magnitude of the electric field of the
crystallographic plane at the particle location point at time
moment $t$ (let us recall that in the case under consideration the
direction of this field is practically orthogonal to the particle
velocity).

 From (\ref{0.2}) it follows that during motion in a bent crystal
 the particle spin rotates about the direction of the momentum
 \cite{VG79,n3}
and owing to the large magnitude of field $E$ ($E \sim 10^7 -
10^8$ CGSE), making the particle following the bending of the
crystal, the frequencies $\Omega_y$ for $(g-2) \sim 1$ are large
$\Omega_y \approx 10^{11} - 10^{13}$ s$^{-1}$ and the angle of
rotation by one centimeter could reach values as large as
$10-10^{3}$ rad/cm.

 An important relation between the spin rotation angle $\theta_s$
and the momentum rotation angle $\theta_m$ in case of planar
motion follows from (\ref{0.3})
\begin{eqnarray}
\theta_s & = & \int_{0}^{T}
\Omega_y(t^{\prime})dt^{\prime}=\left(\frac{g-2}{2}+\frac{1}{\gamma}\right)
\gamma \theta_m, \nonumber \\
\theta_m & = & \int_{0}^{T}\Omega_{0y}(t^{\prime})dt^{\prime}
 \label{0.4}
\end{eqnarray}
Let, for example, the curvature radius of a bent crystal be $10^2$
cm , then for $\gamma=10^2$ and $(g-2) \sim 1$, $\theta_s=1$
rad/cm, that is obviously observable.

 With the help of (\ref{0.4}) we can measure the magnitudes
 $(g-2)$ for a particle without making use concrete models,
 which describe the distribution of intracrystalline fields. It
 is sufficient to measure $\theta_m$ and $\theta_s$:
\begin{eqnarray}
g-2=\frac{2}{\gamma}\frac{\theta_s-\theta_m}{\theta_m}
 \label{0.5}
\end{eqnarray}
If $(g-2) \sim 1$ and $\gamma \gg 1$, then $(g-2) \approx 2
\theta_s/\gamma \theta_m$.

\section{Spin rotation of relativistic particles in the presence of quadrupole interactions
and birefringence effect in pseudoelectric and electric fields}

If a particle has the spin $S \ge 1$ then interaction $\hat{W}_Q$
of its electric quadrupole moment with an inhomogeneous electric
field of crystallographic planes (axis) \cite{n5,n6} and the
birefringence effect in pseudoelectric nuclear and electric fields
could appear important \cite{1}-\cite{me}.

Remember also that a particle with the spin $S \ge 1$ possesses
the electric and magnetic polarizabilities.

Let an  electric field $\vec{E} (\vec{r})$ acts on a particle
(nucleus) ($\vec{r}$ is the particle coordinate).
The energy $\hat{V}_{\vec{E}}$ of interaction of the particle with
the electric field due to the electric polarizability tensor can
be written as \cite{LANL_nastya}:
\begin{equation}
\hat{V}_{\vec{E}}=-\frac{1}{2}\hat{\alpha}_{ik}(E_{eff})_{i}(E_{eff})_{k},
 \label{VE}
\end{equation}
where $\hat{\alpha}_{ik}$ is the electric polarizability tensor of
the particle  , $\vec{E}_{eff}=(\vec{E}+\vec{\beta} \times
\vec{B})$ is the effective electric field; the expression
(\ref{VE}) can be rewritten as follows:
\begin{eqnarray}
\hat{V}_{\vec{E}} =
\alpha_{S}E^{2}_{eff}-\alpha_{T}E^{2}_{eff}\left(\vec{S}\vec{n}_{E}\right)^{2},~
\vec{n}_{E}  =
\frac{\vec{E}+\vec{\beta}\times\vec{B}}{|\vec{E}+\vec{\beta}\times\vec{B}|}
 \label{VE1}
\end{eqnarray}
where $\alpha_{S}$ is the scalar electric polarizability and
$\alpha_{T}$ is the tensor electric polarizability of the
particle.

A particle with the spin $S \ge 1$ also has the magnetic
polarizability, which is described by the magnetic polarizability
tensor $\hat{\beta}_{ik}$ and interaction of the particle with the
magnetic field due to the magnetic polarizability tensor is as
follows \cite{me}:
\begin{equation}
\hat{V}_{\vec{B}}=-\frac{1}{2}\hat{\beta}_{ik}(B_{eff})_{i}(B_{eff})_{k},
 \label{VB}
\end{equation}
where $(B_{eff})_{i}$ are the components of the effective magnetic
field $\vec{B}_{eff}=(\vec{B}-\vec{\beta} \times \vec{E})$;
$\hat{V}_{\vec{B}}$ can be expressed as:
\begin{equation}
\hat{V}_{\vec{B}}=\beta_{S}B_{eff}^{2}-\beta_{T}B_{eff}^{2}\left(\vec{S}\vec{n}_{B}\right)^{2},~
\vec{n}_B=\frac{\vec{B}-\vec{\beta} \times
\vec{E}}{|\vec{B}-\vec{\beta} \times \vec{E}|}.
 \label{VB1}
\end{equation}
where $\beta_{S}$ is the scalar magnetic polarizability and
$\beta_{T}$ is the tensor magnetic polarizability of the particle.

%
When a particle with the spin $S\geq1$ passes through an
unpolarized medium, the medium refraction index depends on
particle spin orientation to its momentum \cite{1,2}.
As a result, for a particle moving in a crystal with the
nonpolarized nuclei the effective potential energy
$\hat{U}^{nucl}$ depends on the spin orientation \cite{1,2,me}
\begin{equation}
\hat{U}^{nucl}=-\frac{2\pi \hbar^{2}}{M \gamma}N (\vec{\rho})
\hat{f(0)},
 \label{1.1}
\end{equation}

 Substituting $\hat{f(0)}$ in its explicit form one can obtain
 \cite{1,2}:
\begin{equation}
\hat{U}^{nucl}=-\frac{2\pi \hbar^{2}}{M \gamma}N (\vec{\rho})
\left(d+d_{1}\left(\vec{S}\vec{n}\right)^{2}\right), \label{1.2}
\end{equation}
 where  $\vec{n}$ is the unit vector along the particle momentum
direction, for $S > 1$ the summands with higher degrees of $S$
could present in (\ref{1.2}) (they are omitted here for
simplicity, so strictly speaking (\ref{1.2}) is applicable for
$S=1$ and $S=\frac{3}{2}$).

Let the quantization axis z is directed along $\vec{n}$ and $m$
denotes the magnetic quantum number. Then, for a particle in the
state that is an eigenstate of the operator $S_{z}$ of spin
projection onto the z-axis, the efficient potential energy can be
written as:
\begin{equation}
\hat{U}^{nucl}=-\frac{2\pi \hbar^{2}}{M \gamma}N (\vec{\rho})
\left(d+d_{1}m^{2}\right). \label{1.3}
\end{equation}

 According to (\ref{1.3}) splitting of the particle energy levels
in a matter is similar to splitting of atom energy levels in an
electric field aroused by the quadratic Stark effect.
 Therefore,
the above effect could be considered as caused by splitting of the
spin levels of the particle in the pseudoelectric nuclear field of
a matter.
%
Rotation and oscillations of deuteron vector and tensor
polarization in matter, along with the spin dichroism
(birefringence effect), appear due to the above interaction.
Spin dichroism was observed in the experiment \cite{4}.

 Thus, considering evolution of the spin of a particle in a
crystal one should take into account several interactions:

1. interactions of the magnetic dipole moment with the
electromagnetic field of a bent crystal;

2. interaction (\ref{VE}) of the particle with the crystal due to
the tensor electric polarizability;

3. interaction of the particle with the crystal due to the tensor
magnetic polarizability;

4. interaction (\ref{1.2}) of the particle with the pseudoelectric
nuclear field of matter;

5. interaction $\hat{W}_{Q}$ of the quadrupole moment with the
inhomogeneous crystallographic electric field.

Therefore, the equation for the particle spin wavefunction is:
\begin{equation}
i\hbar\frac{\partial\Psi(t)}{\partial
t}=\left(\hat{H}_{0}+\hat{W}_{Q}+\hat{U}_{nucl}+\hat{V}_{E}+\hat{V}_{B}\right)\Psi(t)
\label{1.6}
\end{equation}
where $\Psi(t)$ is the particle spin wavefunction,
{$\hat{H}_{0}$ is the Hamiltonian describing the spin behavior
caused by interaction of the magnetic moment with the
electromagnetic field (equation (\ref{1.6}) with the only
$\hat{H}_{0}$  summand converts to the Bargman-Myshel-Telegdy
equation)}.

It is important that the terms $\hat{W}_{Q}$, $\hat{U}_{nucl}$,
$\hat{V}_{E}$ and $\hat{V}_{B}$ cause oscillations and rotation of
spin even in a straight (non-bent) crystal.

Multiple scattering and depolarization can be substantial for
processes in crystals and
{complete description of these processes can be done by the
density matrix formalism (see \cite{LANL_nastya,PR+}}).

So, when a particle moves in a bent crystal the effects of spin
rotation and vector-to-tensor (tensor-to-vector) polarization
conversion appear due to the interactions $\hat{W}_{Q}$,
$\hat{U}_{nucl}$, $\hat{V}_{E}$ and $\hat{V}_{B}$ along with the
spin rotation caused by $(g-2)$. These effects can be used for
measurement $Q$, $\alpha_T$, $\beta_T$ and $d_1$.
 It is of the essence that the above quantities can be measured
 even in straight crystal, but bent crystals could provide higher
 electric fields $E$.


One more additional consequence \cite{Luboshits} follows from
(\ref{0.1}): if the gyromagnetic ratio satisfies the condition $1
< g < 2$, then the electric field does not influence on the spin
of a particle with the energy $\epsilon_0=\frac{g}{g-2}mc^2$ (the
angular velocity of the spin precession appears equal to 0,
whereas the magnetic moment is nonzero). This effect is purely
relativistic and caused by cancellation of ''dynamical''
precession by Thomson precession. When the energy of the particle
is less then $\epsilon_0$, the spin rotates in the same direction
as the momentum does. Otherwise (when the energy of the particle
is greater then $\epsilon_0$), the spin and momentum rotate in
different directions.
The requirement $1 < g < 2$ can be fulfilled for a deuteron
($g=1.72$ and $\epsilon_0=11.5$ GeV) and some nuclei ($^6$Li has
g=1.64).
 In such conditions spin rotation and oscillations appears caused
 only by interactions $\hat{W}_{Q}$,
$\hat{U}_{nucl}$, $\hat{V}_{E}$ and $\hat{V}_{B}$.

%
Let us consider the possibility to measure the tensor electric and
magnetic polarizabilities  $\alpha_T$, $\beta_T$.
 The typical frequency of spin rotation (oscillations) caused by
 the considered interaction due to the particle electric polarizability is:
\begin{equation}
\omega_{\alpha}=\frac{\alpha_T E^2}{\hbar}, \label{5.1}
\end{equation}
the corresponding rotation angle (phase of oscillations) is:
\begin{equation}
\varphi=\omega_{\alpha} \frac{L}{v}, \label{5.2}
\end{equation}
where $L$ is the length of the particle pass inside the crystal,
therefore:
\begin{equation}
\varphi=\frac{\alpha_T E^2 L}{\hbar v}. \label{5.3}
\end{equation}
Hence, $\varphi$ measurement implies $\alpha_T$ measurement.
The angle of momentum rotation in a bent crystal is:
\begin{equation}
\theta_{0}=\frac{e E L}{M c^2 \gamma}, \label{5.4}
\end{equation}
from here
\begin{equation}
E=\frac{M c^2 \gamma}{e L}\theta_{0}, \label{5.5}
\end{equation}
therefore
\begin{equation}
\varphi=\frac{\alpha_T \gamma^2}{\lambda_c r_0 L}{\theta_{0}}^2,
\label{5.6}
\end{equation}
here $\lambda_c$ is the Compton wavelength, $r_0=\frac{e^2}{M
c^2}$ is the electromagnetic radius of the particle.
 And finally, measuring $\varphi$, $\theta_{0}$, $L$ and $\gamma$
 we can find $\alpha_T$:
 \begin{equation}
\alpha_T=\frac {\lambda_c r_0 L} {\gamma^2}
\frac{\varphi}{{\theta_{0}}^2}, \label{5.7}
\end{equation}

Let us evaluate $\alpha_T$ could be measured when a particle
passes through a crystal. Suppose that the experimentally measured
angle of rotation is about $\varphi \approx 10^{-4}$.
 The strength
of the electric field can be achieved in a bent crystal is about
$E \approx 10^9$ CGSE.
 Therefore we can obtain the estimation
 $\alpha_T=\frac {\hbar c \varphi} {E^2 L}=\frac{3}{L}10^{-39}$
 cm$^3$. The electric polarizability of a deuteron can be evaluated as $\alpha_T \approx
 10^{-40}$ cm$^3$ \cite{theory_betaT}. The similar estimations can be obtained for $\beta_T$ \cite{theory_betaT}.

 Then, it seems possible to measure of the tensor polarizability of
 deuterons (nuclei)  and as well as to get limits
 for polarizability of elementary particles (for example, $\Omega$
 hyperon).

\section{Particle motion in a polarized crystal}

Suppose now that the crystal nuclei are polarized.
The cross-section of high-energy particle scattering  by a nucleus
(the amplitude of zero-angle scattering) in a crystal with the
polarized nuclei depends on the particle spin orientation with
respect to target polarization.
Therefore, the absorption coefficient also depends on particle
spin orientation.
Let a particle moves in a crystal at a small angle to the
crystallographic plane (axis) close to the Lindhard angle.
In this case the
 coefficient of particle absorption by nuclei appears greater than the
 absorption coefficient for the particle moving in the crystal at a large angle to the
crystallographic plane (axis).
 Let us discuss the possibility to use a polarized crystal to obtain a polarized beam of high energy
 particles and to analyze the polarization state of high energy particles \cite{Vesti}.

 In the case of consideration a positively charged high-energy particle moves close to the
 top of the potential barrier, therefore, it moves in the range with the nuclei
 density higher than that for an amorphous medium.
If it moves at a small angle to the crystallographic plane, then
the growth is described by
  $\frac{a}{a_0} \approx
 10^2$, here $a$ is the lattice period, $a_0$
 is the amplitude of the nucleus thermal oscillations.
If it moves at a small angle to the crystallographic axis, then
the growth is described by $(\frac{a}{a_0})^2 \approx
 10^4$.

 The polarization degree after passing the polarized crystal with the
 thickness $L$ is
 \begin{equation}
 P=\frac{e^{-\rho \sigma_{\uparrow \uparrow}L}-e^{-\rho \sigma_{\uparrow \downarrow}L}}
 {e^{-\rho \sigma_{\uparrow \uparrow}L}+e^{-\rho \sigma_{\uparrow \downarrow}L}}
 \end{equation}
where $\sigma_{\uparrow \uparrow}$ is the total scattering
cross-section for the particle with the spin parallel to the
polarization vector of nuclei and $\sigma_{\uparrow \downarrow}$
is the total scattering cross-section for the particle with the
spin anti-parallel to the polarization vector of nuclei.
The effect magnitude is determined by the parameter
$A=\rho(\sigma_{\uparrow \uparrow}-\sigma_{\uparrow \downarrow})L$
($A=\rho_{av} \cdot 10^2 \Delta \sigma L$ for a plane and
$A=\rho_{av} \cdot 10^4 \Delta \sigma L$ for an axis, $\rho_{av}$
is the average density of nuclei in the crystal).

The length $L_1$ corresponding to $A=1$ is equal to $L_1 =
\frac{1}{\rho \Delta \sigma}$. If suppose $\rho_{av} \approx
10^{22}$, $\Delta \sigma \sim 10^{-25}~cm^2$ then for a plane $L_1
\approx 10 ~ cm$; for an axis $L_1 \approx 10^{-1} ~ cm$

Let positively charged particles move parallel to a
crystallographic plane then the most of them appear close to the
bottom of the potential well and far from nuclei, therefore, the
absorption coefficient is smaller comparing that for average
crystal absorption.

In contrast to the positively charged particles a negatively
charged particle (for example $\Omega^{-}$ hyperon, antiproton),
being channelled, moves in the range with high density of
 nuclei,therefore, even a very thin polarized crystal ($L \approx 10^{-1}$ cm)
  can
be an effective polarizer and polarization analyzer.

Moreover, for a negatively charged channelled particle the angle
of spin rotation in the nuclear pseudomagnetic field of the
polarized crystal \cite{1,2,PR+} can be larger comparing with that
for amorphous matter (for example, for the particle
 energy $\sim 10~GeV$ in channelling conditions in the
crystal of $\approx 10^{-1}$ cm length the angle of spin rotation
is about $\vartheta \sim 10^{-1}$ rad).
Similarly the birefringence effect in pseudoelectric nuclear field
and electric field also grows  for $\Omega^{-}$ hyperon channelled
in a crystal (or for any other channelled particle with the
negative charge).
This is the reason for the thin polarized crystals to be used as
nuclear-optical elements can guide polarization of high-energy
particle beams.

To increase interaction of a positively charged particle with a
nucleus a bent crystal could be used.
In this case the centrifugal forces push the channelled beam to
the range with the high density of nuclei.

Thus, both the straight and bent crystals provide methods for
getting the polarized beams of high-energy particles, rotating the
particle spin and analyzing particle polarization state.

Note that spin-orbit interaction also leads to particle
polarization as well as to the left-right asymmetry in polarized
particles scattering.
The density of nuclei near the axis is much higher than the
density of amorphous target. This makes the induced polarization
(left-right asymmetry) for particles scattered by the crystal axes
noticeably higher, as compared with amorphous matter, due to
interference of the Coulomb and effective nuclear potentials of
the axis \cite{Vesti}.

\section{Conclusion}
Thus, particle motion in a bent (straight) crystal is accompanied
by particle spin rotation and oscillations that allows to measure
the tensor electric polarizability of nuclei and elementary
particles.

Channelling of particles in either straight or bent crystal with
the polarized nuclei could be used for polarization or analyzing
polarization of high energy particles. The beam of nonpolarized
particles extracted from the storage ring could be significantly
polarized by applying the additional polarized bent (straight)
crystal.

This work is prepared in the framework of INTAS Project$\#$
03-52-6155.
%


\end{document}